\documentclass[prb,fleqn,12pt,showpacs]{revtex4-1}
\usepackage{epsfig}
\usepackage{graphicx}
\usepackage{amsfonts,amssymb}
\usepackage{amsmath}
\usepackage{color}
\definecolor{refcolor}{rgb}{0.0,0.0,1.0} 

\newcommand{\be}{\begin{equation}}
\newcommand{\ee}{\end{equation}}   
\newcommand{\bea}{\begin{eqnarray}}
\newcommand{\eea}{\end{eqnarray}}
\newcommand{\ba}{\begin{array}}
\newcommand{\ea}{\end{array}}
\newcommand{\phr}[1]{Phys.~Rev. {\bf #1}}
\newcommand{\phrl}[1]{Phys.~Rev.~Lett. {\bf #1}}
\newcommand{\phrb}[1]{Phys.~Rev.~B {\bf #1}}

\renewcommand{\k}{{\bf k}}
\newcommand{\Q}{{\bf Q}}

\begin{document}
\title{Charge order induced in an orbital density-wave state}
\author{Dheeraj Kumar Singh$^1$,$^2$}
\author{Tetsuya Takimoto$^1$}
\email{takimoto@hanyang.ac.kr} 
 
\affiliation{$^1$Department of Physics, Hanyang University, 
17 Haengdang, Seongdong, Seoul 133-791, Korea}
\affiliation{$^2$Asia Pacific Center for Theoretical Physics, Pohang, Gyeongbuk 790-784, Korea}

\begin{abstract}
Motivated by the recent ARPES measurements (Evtushinsky et. al., PRL {\bf 105}, 147201 (2010)) and evidence for the density-wave state for the charge and 
orbital ordering (Garc\'{i}a et al., PRL {\bf 109}, 107202 (2012)) in La$_{0.5}$Sr$_{1.5}$MnO$_4$, the issue of charge and orbital ordering in a two orbital tight-binding model for layered manganite near half doping is 
revisited. We find that the charge order with an ordering wavevector 2$\Q$ = $(\pi, \pi)$ is induced by the orbital order of $B_{1g}$ representation with a different ordering wavevector $\Q$, where the primary order 
parameter results from the strong Fermi-surface nesting. The orbital and charge order parameters develop according to $\sqrt{T_{CO}-T}$ and $T_{CO}-T$, respectively, by decreasing the temperature below the orbital
ordering temperature $T_{CO}$.  Moreover, the orbital order is found to stabilize the CE-type spin arrangement observed experimentally below $T_{CE} < T_{CO}$. 
\end{abstract}

\maketitle
\newpage
\section{Introduction}
Orbital degree of freedom plays a crucial role not only in the transport properties as revealed in several experiments near metal-insulator transition 
associated with the colossal magnetoresistance (CMR),\cite{saitoh,moussa_2007} but also in stabilizing 
a range of magnetically ordered phases in manganites. For instance, the role of orbital 
ordering was recognized in stabilizing CE-type spin arrangement very early.\cite{goodenough} This state consists 
of ferromagnetically ordered zigzag chains with spins on 
the neighboring chains oriented in the opposite directions. In addition, the 
state also exhibits the charge and orbital ordering with wavevectors 
($\pi, \pi, 0$) and ($0.5\pi, 0.5\pi, 0$), respectively, with more charge accumulated at 
the orbitally polarized sites.\cite{wollan,moritomo,zheng} Several theoretical 
investigations have supported $d_{3x^2-r^2}$/$d_{3y^2-r^2}$-type orbital order, in which $d_{3x^2-r^2}$ and $d_{3y^2-r^2}$
orbitals are occupied at the bridge sites of the parts of the zigzag chain that run
parallel to the $x$ and $y$ directions, respectively, so 
that the kinetic energy gain through the double 
exchange mechanism lowers the energy of the 
system with respect to any other possible orbital order.\cite{yunoki,hotta,sboychakov} However, there is no 
consensus among various x-ray experiments regarding
whether $d_{3x^2-r^2}$/$d_{3y^2-r^2}$-\cite{dhesi,wu} or $d_{x^2-z^2}$/$d_{y^2-z^2}$\cite{huang,wilkins}-type orbital order exists. 
 
Half-doped single-layer manganite La$_{0.5}$Sr$_{1.5}$MnO$_4$, apart from showing the CE-type ordered 
state with a transition temperature  $T_{CE}$ $\approx$ 110 $K$, also exhibits a charge and orbitally ordered state (CO) with the transition temperature
$T_{CO} \approx$ 220 $K$ above $T_{CE}$.\cite{moritomo} The CO state has been observed in different single-layer manganites in the electron-doped region by several 
experiments such as  x-ray experiments,\cite{Larochelle,nagai} optical 
spectroscopy,\cite{lee} and high-resolution electron microscopy (HREM).\cite{moritomo_jpsj} Some of these experiments have described this state as a charge-density wave of $d_{x^2-y^2} (d_{3z^2-r^2})$ electrons,
where orbital ordering wavevector ${\bf Q}$ is related
to the hole doping $x$ through ${Q}_x = {Q}_y = \pi(1-x)$.\cite{Larochelle} Moreover, incommensurate sinusoidal modulations have been observed in a recent experiment.\cite{garcia} The stabilization of density-wave state may imply 
a very important role of the structure of the Fermi surface not only for the CO state but also for the CE-type AFM state.

A density-wave state scenario\cite{yao} for the CO state has been further anticipated by a recent 
angle-resolved photoemission spectroscopy (ARPES) 
on the half-doped La$_{0.5}$Sr$_{1.5}$MnO$_4$.\cite{{evtushinsky}} The Fermi surface comprises of
a small circular electron pocket around the $\Gamma$ point and a relatively 
large hole pocket around the M point. Major portions of the hole pockets 
are largely flat, and therefore susceptible to the nesting instability. Furthermore,
the nesting vector (0.5$\pi$, 0.5$\pi$) is also in agreement with the orbital ordering wavevector.  

Several studies have carried out to clarify the origin of the charge order. For instance in the case of CE state, it has been argued 
that the inter-orbital interaction can also lead to the 
charge ordering. This may follow from the fact that an electron at a corner site with nearly equal mixture of both $d_{x^2-y^2}$ and $d_{3z^2-r^2}$ orbitals feels strong interorbital interaction as compared to an electron 
at the bridge site with near-complete orbital polarization, and thereby is 
pushed to the neighboring bride sites to lower the energy of the system.\cite{brink} Other studies
have emphasized on the role of long-range Coulomb interaction
in stabilizing the 
charge order.\cite{yin,volja} including the density-wave state scenario where orbital order originates from the Fermi surface nesting,\cite{yao} though the difficulty with it is two fold.
First, it requires a fine tuning of the interaction parameter  
to describe the composite appearance of charge and 
orbital order at $T_{CO}$. Secondly, the form of interaction 
necessary to explain the charge ordering away from half doping in the density-wave state with charge ordering wavevector 2${\bf Q}$ will be unrealistic. 

In this paper, we revisit the issue of orbital ordering in La$_{0.5}$Sr$_{1.5}$MnO$_4$ for both the CO state without any
spin order and with CE-type AFM order by considering a two orbital tight-binding model which includes the salient features of the recent 
experimental Fermi surface. While the orbital order may result from the Fermi surface nesting, the origin of a simultaneous appearance of charge order is 
investigated within the meanfield description employing the Jahn-teller phononic approach. Our investigation also attempts to explain if
the charge and orbital ordering transition accompanies a metal-insulator transition especially at half doping. The behavior of order parameters as a function of temperature is also compared with experimental data of the x-ray.
\section{Model Hamiltonian}
We consider a two orbital Hamiltonian for the single-layer manganites in the basis of $d_{x^2-y^2}$ and $d_{3z^2-r^2}$ orbitals
\begin{eqnarray}
{\mathcal H} &=& -\sum_{{\bf i} \gamma \gamma' \sigma  {a}}
  t^{\bf a}_{\gamma \gamma'} d_{{\bf i} \gamma \sigma {}}^{\dag}
  d_{ {\bf i+a} \gamma' \sigma }   +\varepsilon_z \sum_{\bf i}\mathcal{T}^z_{\bf i}+ \sum_{ {\bf i} l}
 gq_{{\bf i}l} \mathcal{T}^l_{{\bf i}}+\sum_{ {\bf i} l} \mathcal{K}_l q^2_{{\bf i} l }/2
  -2J_{\rm H} \sum_{{\bf i}} {\bf S}_{{\bf i}} \cdot {\bf s}_{{\bf i}}.
\end{eqnarray}
The kinetic term within the tight-binding description
includes $d^{\dagger}_{{\bf i} 1 \sigma }$ ($d^{\dagger}_{{\bf i} 2 \sigma}$) as
the electron creation operator at site ${\bf i}$ with spin $\sigma$ in the orbital $d_{x^2-y^2}$ ($d_{3z^2-r^2}$). 
$t^{\bf a}_{\gamma \gamma'}$ are the
 hopping elements between $\gamma$ and $\gamma'$ orbitals along ${\bf a}$ 
connecting the nearest-neighboring sites, 
which are given by
$t^{\bf x}_{11}$ = $-\sqrt{3}t^{\bf x}_{12}$ = $-\sqrt{3}t^{\bf x}_{21}$ = $3t^{\bf x}_{22}$ = $3t/4$ 
for ${\bf a}$ = ${\bf x}$ and
$t^{\bf y}_{11}$ = $\sqrt{3}t^{\bf y}_{12}$ = $\sqrt{3}t^{\bf y}_{21}$ = $3t^{\bf y}_{22}$ = $3t/4$
for ${\bf a}$ = ${\bf y}$, respectively. $t$ is set to be the unit of energy in the following.
Second term accounts for the crystalline-electric field (CEF) responsible for the splitting of $e_g$ levels in 
the tetragonal symmetry, where $\mathcal{T}^z_{\bf i} = \sum_{\sigma} \psi^{\dagger}_{{\bf i}\sigma } \hat{\tau}^z \psi_{ {\bf i}\sigma} $ 
with $\psi^{\dagger}_{{\bf i}\sigma } = (d^{\dagger}_{{\bf i}1 \sigma }, d^{\dagger}_{{\bf i} 2 \sigma })$ and $\hat{\tau}^z$ is the $z$-component of the 
Pauli matrices in the orbital space. According to the convention 
adopted here, a positive $\varepsilon_z$ which favors the occupancy of $d_{3z^2-r^2}$
orbital over $d_{x^2-y^2}$ orbital. Third term describes the coupling between the electron and Jahn-Teller distortions, where  $q_{ {\bf i} x}$ and $q_{ {\bf i}z}$ 
correspond to transverse ($x^2-y^2$)- and longitudinal ($3z^2-r^2$)-type Jahn-Teller distortions, respectively, and $\mathcal{T}^x_{\bf i} = \sum_{\sigma} \psi^{\dagger}_{{\bf i}\sigma }
\hat{\tau}^x \psi_{{\bf i} \sigma } $ with $\hat{\tau}^x$ as the $x$-component of the 
Pauli matrices. Fourth term accounts for the potential energy of these distortions with 
$\mathcal{K}_l$ as the spring constant. Fifth term represents the 
Hund's coupling ($J_{\rm H}$) between the spin ${\bf s}_i=\sum_{\gamma \sigma \sigma'} d_{ {\bf i} \gamma \sigma 
}^{\dag} {\pmb{\sigma}}_{\sigma \sigma'} d_{ {\bf i} \gamma \sigma' }$ of $e_g$ electrons and the
 localized $t_{2g}$ spin ${\bf S}_i$. The intra- and inter-orbital Coulomb interactions have not been considered here as 
 their inclusion do not change the essential physics as reported earlier.\cite{hotta} This simplified model has been extensively studied using Monte Carlo
 simulation to study the various orderings in manganites.\cite{yunoki,hotta1}
  \begin{figure}
\begin{center}
\vspace*{-2mm}
\hspace*{0mm}
\psfig{figure=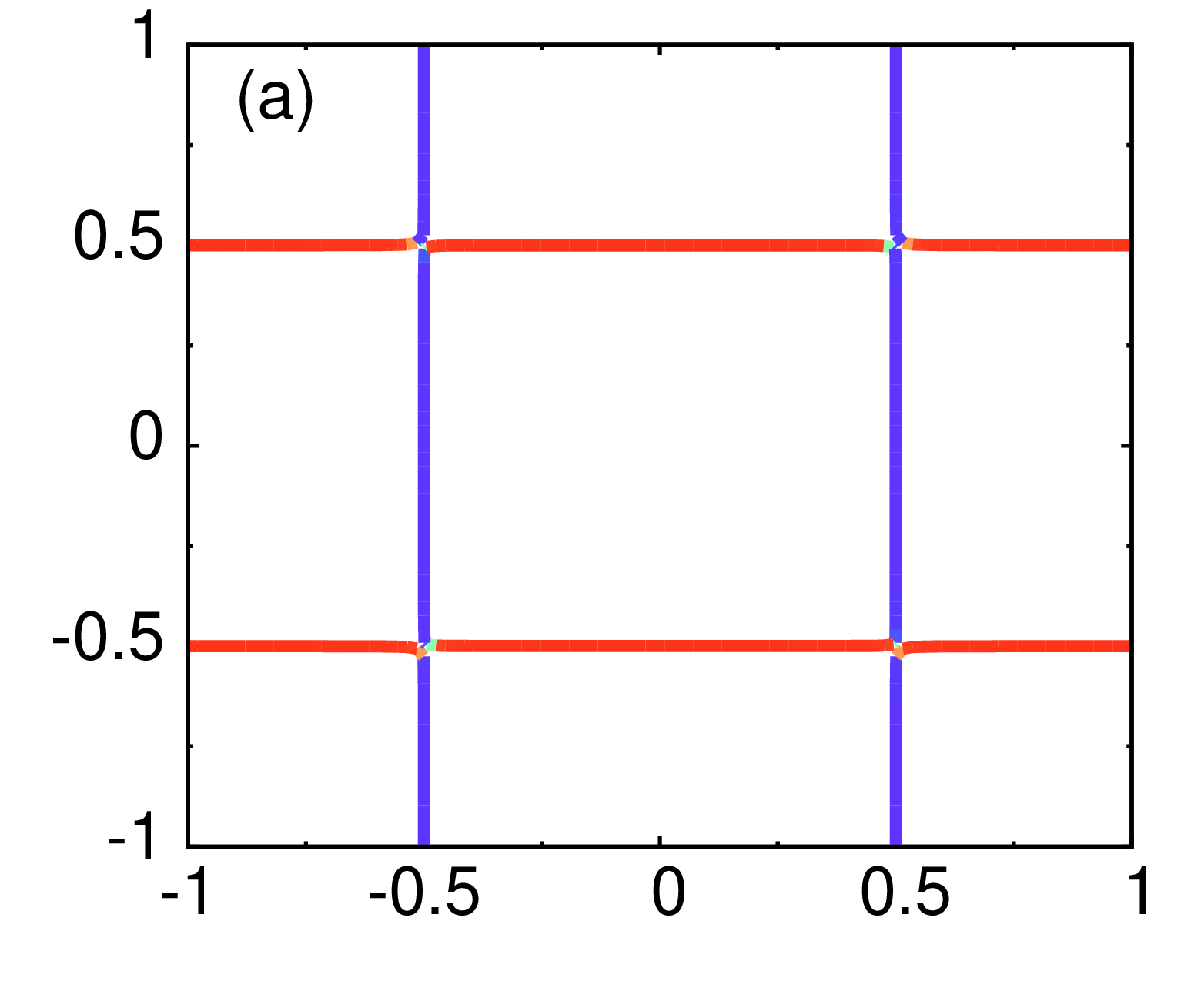,width=42mm,angle=-90}
\psfig{figure=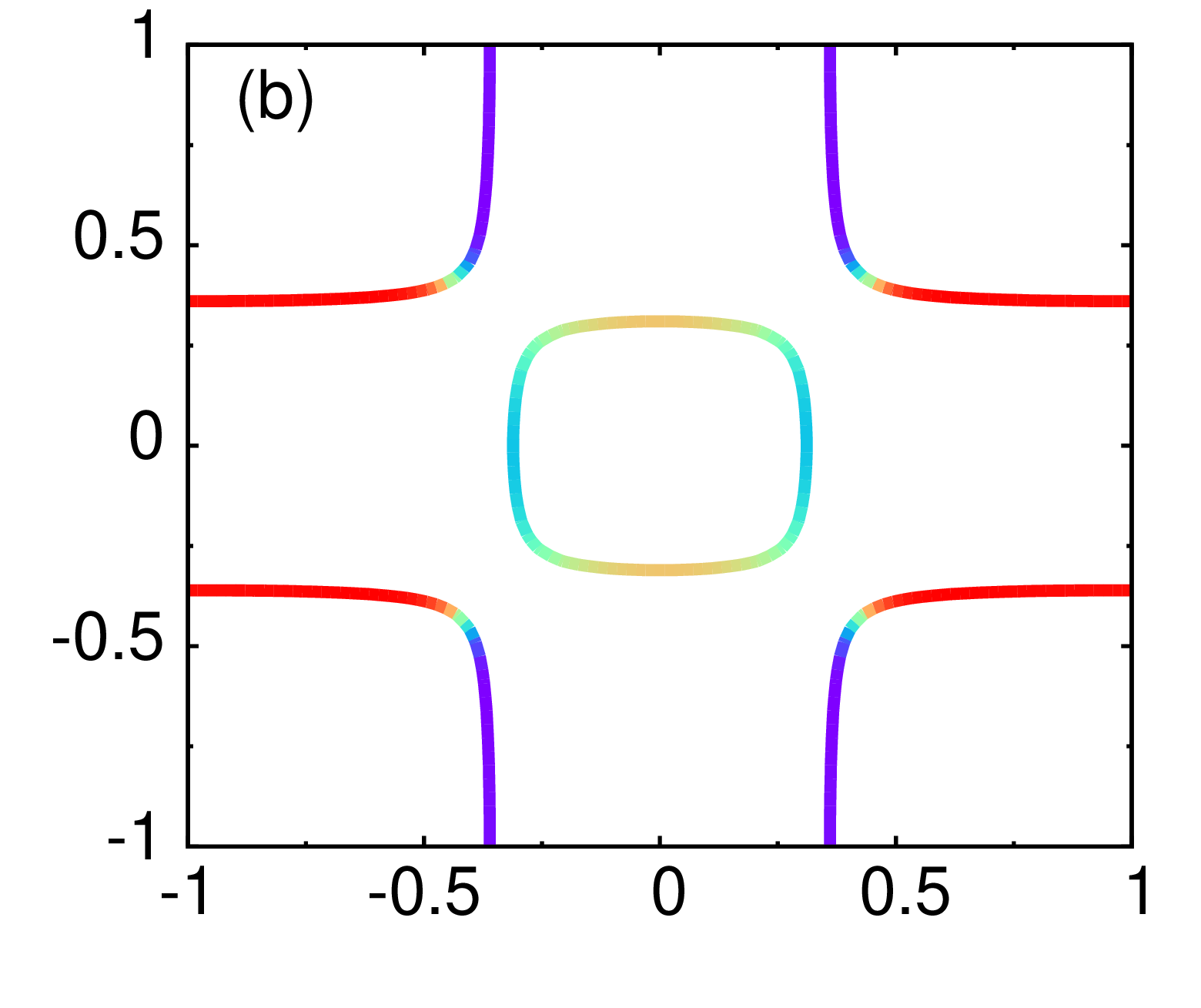,width=42mm,angle=-90}
\psfig{figure=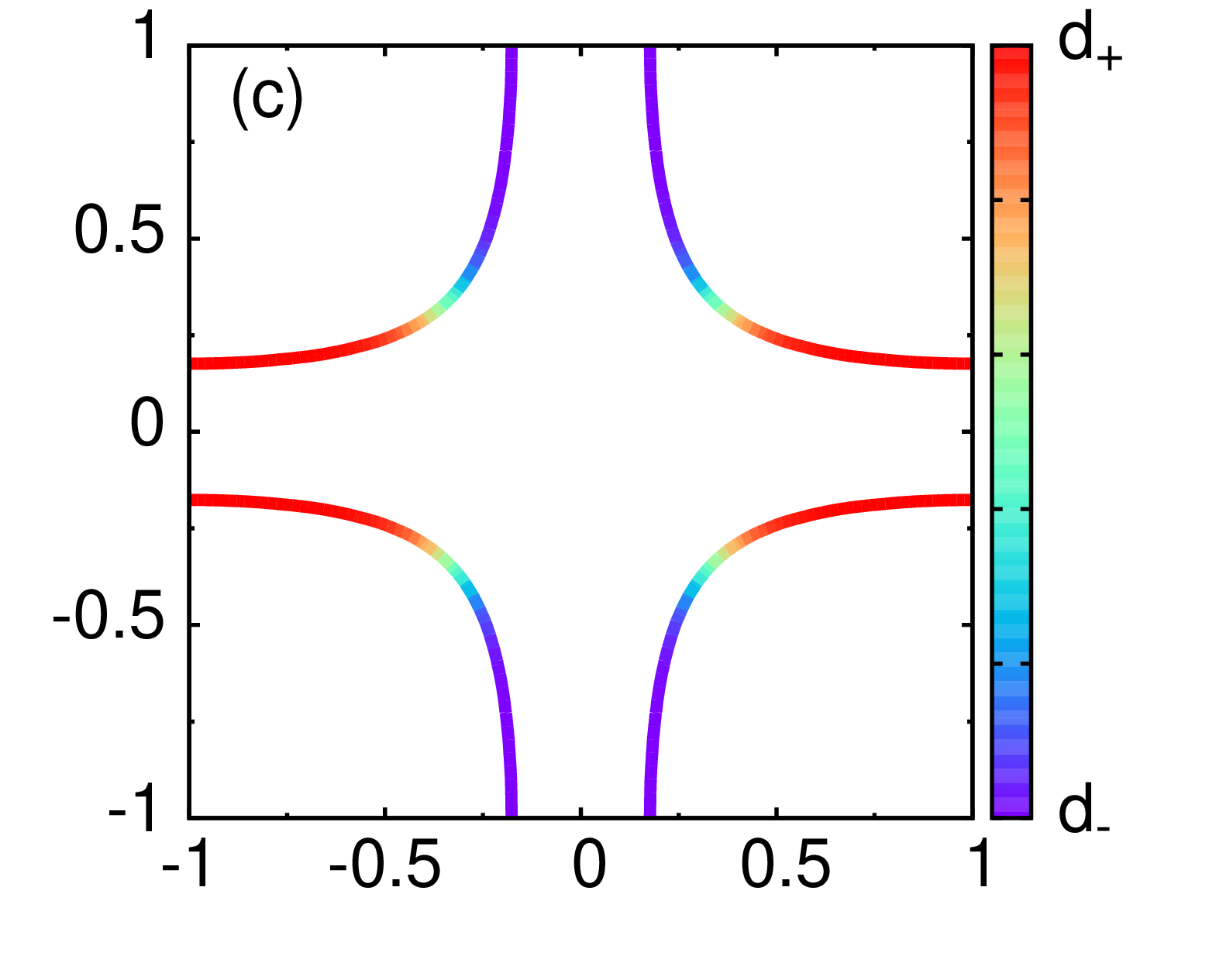,width=42mm,angle=-90}
\vspace*{-5mm}
\end{center}
\caption{Orbital density distributions on the Fermi surface using $d_{\mp}$ basis for different hole dopings (a) $x$ = 0, (b) $x$ = 0.3, and (c) $x$ = 0.6, where 
$\varepsilon_z$ = 0.}
\label{fs}
\end{figure} 

Fig. \ref{fs} shows the orbital densities on the Fermi surface obtained from kinetic part of the Hamiltonian for several values of hole doping $x$ in the orbital basis 
of $d_-$ and $d_+$ given by $d_{\mp} = \frac{1}{\sqrt{2}}(d_{3x^2-r^2}\mp d_{x^2-y^2}$). There exists a good nesting for all hole dopings with strongest being at $x = 0$,\cite{efremov} and nesting 
vector components are approximately given by $Q_x = Q_y = \pi(1 - x)$. A common feature of the Fermi surfaces for these dopings is the near-complete orbital polarization of the flat regions of the hole pockets
susceptible to the Fermi surface nesting, which can lead to the longitudinal orbital instability in the basis consisting of $d_-$ and $d_+$ orbitals. Furthermore, considering symmetry in the tetragonal crystal,
orbital order can be either of $d_{x^2-y^2}/d_{3z^2-r^2}$-type or $d_{-}/d_{+}$-type corresponding to one dimensional
representations $A_{1g}$ or $B_{1g}$, respectively.\cite{dk}   
\section{charge and orbitally ordered state at half doping}
\begin{figure}
\begin{center}
\vspace*{-2mm}
\hspace*{0mm}
\psfig{figure=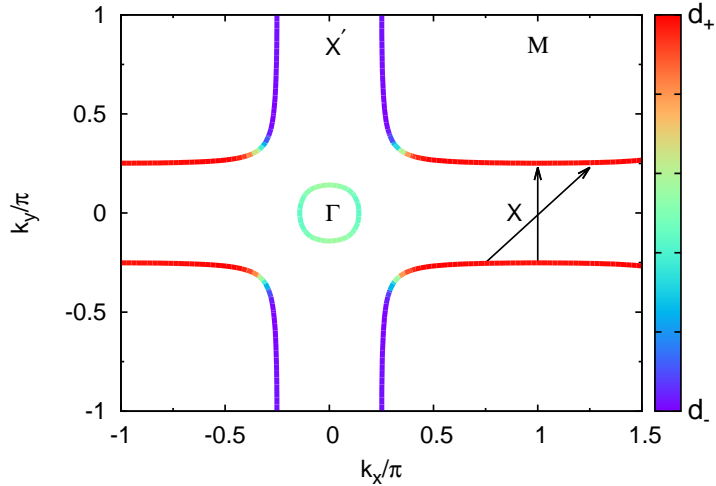,width=70mm,angle=-90}
\vspace*{-5mm}
\end{center}
\caption{Fermi surface plotted in the basis of $d_{-}$ and $d_{+}$ for chemical potential $\mu = -1.25$, and $\varepsilon_z = 0.3$, where $d_{\mp}= \frac{1}{\sqrt{2}}( d_{3z^2-r^2} \mp d_{x^2-y^2})$. Various features including 
a small hole pockets around the $\Gamma$ point is in good agreement with the experiments. \cite{evtushinsky} The Fermi surface is predominantly composed of $d_-$ and $d_{+}$ orbitals around X and X$^{\prime}$, respectively.}
\label{fs_halfdop}
\end{figure} 
Half-doped La$_{0.5}$Sr$_{1.5}$MnO$_4$ exhibits a charge and orbitally ordered state in the temperature range $T_{CE} < T < T_{CO}$, 
with the charge and orbital ordering wavevectors $\Q_c$ = $(\pi, \pi)$ and $\Q_o$ = $(0.5\pi, 0.5\pi)$, respectively. 
To explore the nature of the CO state with the local moment thermally disordered, we consider large 
Hund's coupling limit where the double occupancy of the $e_g$ orbitals is not allowed energetically and the 
$e_g$ spins are enslaved to the $t_{2g}$ spins. Therefore, we drop the spin index in this Section. Then, the Fermi surface corresponding to the electron
density persite 1 - $x$ $\approx$ 0.52 and the crystalline-electric field $\varepsilon_z$ = 0.3 is shown in Fig. \ref{fs_halfdop}, which 
is in good agreement with the experimental ARPES result.\cite{evtushinsky} The Fermi surface, 
consists of a small circular electron pocket around the $\Gamma$ point and a large hole pocket around the M point with portions being 
almost flat near the zone boundary at the X point. Of the two nesting possibilities with vectors (0.5$\pi$, 0) and (0.5$\pi$, 0.5$\pi$), latter one shows the 
stronger orbital ordering instability\cite{dk} and therefore will lead to the 
orbital ordering of $B_{1g}$-representation with wavevector (0.5$\pi$, 0.5$\pi$).

\begin{figure}
\begin{center}
\vspace*{-2mm}
\hspace*{0mm}
\psfig{figure=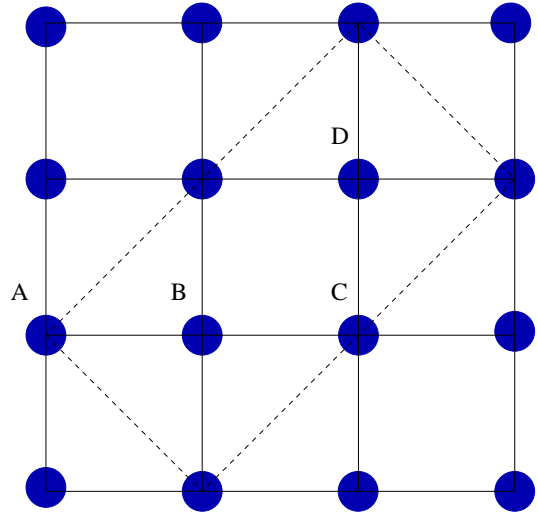,width=70mm,angle=0}
\vspace*{-5mm}
\end{center}
\caption{Unit cell in the charge and orbitally ordered state with the orbital ordering wavevector (0.5$\pi$, 0.5$\pi$) indicated by the dashed line, while $A$, $B$, $C$ and $D$ are the 
sublattices described in the text.}
\label{uc}
\end{figure} 
To describe the orbital density-wave state resulting from the Fermi-surface nesting with ordering wavevector $\Q = (0.5\pi, 0.5\pi)$,
the electron-phonon interaction term by the meanfield decoupling\cite{hotta} can be reduced to 
\be
\hat{H}_{mf} = -  \left[ \frac {\mathcal{J}_x}{2N} (\mathcal{T}^x_{\Q} \langle \mathcal{T}^x_{-\Q} \rangle +
\mathcal{T}^x_{-\Q} \langle \mathcal{T}^x_{\Q} \rangle)+\frac {\mathcal{J}_z}{2N} (\mathcal{T}^z_{\Q} \langle \mathcal{T}^z_{-\Q} \rangle +
\mathcal{T}^z_{-\Q} \langle \mathcal{T}^z_{\Q} \rangle) \right] 
\ee
with  $\mathcal{J}_x = g^2/\mathcal{K}_x$, $\mathcal{J}_z = g^2/\mathcal{K}_z$. Here, it should be noted that $\langle \mathcal{T}^x_{\Q} \rangle$ = $\langle \mathcal{T}^x_{-\Q} \rangle$ is the 
primary order parameter of the orbital order of $B_{1g}$-representation with $\Q = (0.5\pi, 0.5\pi)$ and $\langle \mathcal{T}^z_{\Q} \rangle$ = 0 in the present case.

Normal and anomalous Green's functions in the orbitally ordered state with wavevector $n{\bf Q}$ are defined as
\bea
G_{\gamma \gamma^{\prime}}(\k,\zeta)&=&-\langle T_{\zeta}[c_{\k \gamma}(\zeta)c^{\dagger}_{\k \gamma^{\prime}}(0)], \nonumber\\ 
F^{n\Q}_{\tau \tau^{\prime}}(\k,\zeta)&=&-\langle T_{\zeta}[c_{\k+n\Q \gamma}(\zeta)c^{\dagger}_{\k \gamma^{\prime}}(0)],
\eea
where $n =$ 1, 2 and 3. Then, the charge and orbital ordered state can be described by the following equations of motion
for the Green's functions
\bea
\hat{G}(\k,i\omega_n)&=&\hat{G}^{0}(\k,i\omega_n)-\hat{G}^{0}(\k,i\omega_n) (\frac{\mathcal{J}_x}{2N} \hat{\tau}^x \langle \mathcal{T}^x_{\Q}\rangle + \frac{\mathcal{J}_z}{2N} \hat{\tau}^z \langle \mathcal{T}^z_{\Q}\rangle)(\hat{F}^{\Q}(\k,i\omega_n)+\hat{F}^{3\Q}(\k,i\omega_n)) \nonumber\\ 
\hat{F}^{\Q}(\k,i\omega_n)&=& -\hat{G}^{0}(\k+\Q,i\omega_n) (\frac{\mathcal{J}_x}{2N} \hat{\tau}^x \langle \mathcal{T}^x_{\Q}\rangle + \frac{\mathcal{J}_z}{2N} \hat{\tau}^z \langle \mathcal{T}^z_{\Q}\rangle) (\hat{F}^{2\Q}(\k,i\omega_n)+\hat{G}(\k,i\omega_n))\nonumber\\
\hat{F}^{2\Q}(\k,i\omega_n)&=& -\hat{G}^{0}(\k+2\Q,i\omega_n) (\frac{\mathcal{J}_x}{2N} \hat{\tau}^x \langle \mathcal{T}^x_{\Q}\rangle + \frac{\mathcal{J}_z}{2N} \hat{\tau}^z \langle \mathcal{T}^z_{\Q}\rangle)(\hat{F}^{\Q}(\k,i\omega_n)+\hat{F}^{3\Q}(\k,i\omega_n))\nonumber\\
\hat{F}^{3\Q}(\k,i\omega_n)&=& -\hat{G}^{0}(\k+3\Q,i\omega_n) (\frac{\mathcal{J}_x}{2N} \hat{\tau}^x \langle \mathcal{T}^x_{\Q}\rangle + \frac{\mathcal{J}_z}{2N} \hat{\tau}^z \langle \mathcal{T}^z_{\Q}\rangle) (\hat{F}^{\Q}(\k,i\omega_n)+\hat{G}(\k,i\omega_n)),
\label{green}
\eea
where the single electron Matsubara Green's function is given by
\begin{equation}
\hat{G}^{(0)} ({\bf k}, i \omega_n ) = \frac{\left( i \omega_n - \varepsilon^+_{\bf k} +\mu \right) \hat{\tau}^0 - \varepsilon^-_{\bf k} 
\hat{\tau}^z + \varepsilon^{12}_{\bf k} \hat{\tau}^x}{\left(i \omega_n - E^+_{\bf k} \right) \left(i \omega_n - E^-_{\bf k} \right)}
\end{equation}
with $\hat{\tau}^0$ as a 2 $\times$ 2 identity matrix and Fermionic Matsubara frequency $\omega_n=(2n+1)\pi T$. Also
\begin{eqnarray}
  \varepsilon^+_{\bf k} & = & \frac{1}{2}(\varepsilon^{11}_{\bf k}+\varepsilon^{22}_{\bf k}), \,\,\, \varepsilon^-_{\bf k}  =  \frac{1}{2}(\varepsilon^{11}_{\bf k}-\varepsilon^{22}_{\bf k})+\varepsilon_z, 
 \end{eqnarray}
and
 \begin{eqnarray}
   \varepsilon^{11}_{\bf k} & = & -\frac{3}{2}(\cos k_x+\cos k_y) \nonumber\\
   \varepsilon^{12}_{\bf k} & = & \frac {\sqrt{3}}{2}(\cos k_x-\cos k_y)\nonumber\\
    \varepsilon^{22}_{\bf k} & = & -\frac{1}{2}(\cos k_x+\cos k_y).
\end{eqnarray} 
The electronic dispersion measured from the chemical potential is obtained as
\begin{equation}
E^{\pm}_{\bf k} = \varepsilon^+_{\bf k} \pm \sqrt{(\varepsilon^-_{\bf k})^2 +
(\varepsilon^{12}_{\bf k})^2} - \mu\label{Ek}.
\end{equation}
Here, we note that $\hat{F}^{3\Q}(\k,i\omega_n) = \hat{F}^{-\Q}(\k,i\omega_n)$ as 4$\Q$ is the reciprocal lattice vector in the original Brillouin zone. Importantly, order parameters with 
wavevector 2$\Q$ are induced either $\langle \mathcal{T}^x_{\Q} \rangle$ or $\langle \mathcal{T}^z_{\Q} \rangle$ is non-zero which can
easily be seen from Eq. \ref{green}. Moreover, the induced order parameter varies as a square of 
the primary order parameter. In fact as we will see below, the transverse orbital order $\langle \mathcal{T}^x_{\Q} \rangle$ induces the charge ordering with wavevector 2$\Q$.

For convenience, we transform 
the Hamiltonian from the momentum basis for the ordered state in the reduced Brillouin zone to the sublattice basis as 
described in the Appendix. Thus, the meanfield Hamiltonian in a four-sublattice basis\cite{as} is given by the following 8$\times$8 matrix form
\bea
\hat{H}_{CO}({\bf k}) &=& \sum_{{\bf k}, \mu}
\Psi^{\dagger}_{\bf k}
\begin{pmatrix}
\hat{\Delta}_{cr} & \hat{K} & \hat{0} & \hat{K}^{\dagger} \\
\hat{K}^{\dagger} & -\hat{\Delta}_o+\hat{\Delta}_{cr} & \hat{K} & \hat{0} \\
\hat{0} & \hat{K}^{\dagger} & \hat{\Delta}_{cr} & \hat{K} \\
\hat{K} & \hat{0} & \hat{K}^{\dagger} & \hat{\Delta}_o+\hat{\Delta}_{cr}
\end{pmatrix}
\Psi_{\bf k},
\label{co}
\eea
where $\Psi^{\dagger}_{\bf k} = (d^{\dagger}_{A {\bf k} 1}, 
d^{\dagger}_{ A {\bf k} 2}, d^{\dagger}_{ B {\bf k} 1}...d^{\dagger}_{ D {\bf k} 1}, d^{\dagger}_{D {\bf k} 2} )$ is the electron field 
operator. $A, B, C$ and $D$ denote the four sublattices as shown in Fig. \ref{uc}. Orbital exchange and crystalline-electronic fields are given by 2$\times$2 matrices
\bea
 \hat{\Delta}_o &=& \mathcal{J}_x m_{ox}\hat{\tau}^x + \mathcal{J}_z m_{oz}\hat{\tau}^z  \nonumber\\
 \hat{\Delta}_{cr} &=& \varepsilon_z \hat{\tau}^z,
\eea
with $m_{ox} = \langle \mathcal{T}^x_{\Q} \rangle$ and $m_{oz} = \langle \mathcal{T}^z_{\Q} \rangle$. The Bloch phase factor is  
\bea
 \hat{K}^{\dagger} =
-\frac{1}{4}
\begin{pmatrix}
3(e^{i k_x}+e^{i k_y}) & -\sqrt{3}(e^{i k_x}-e^{i k_y}) \\
-\sqrt{3}(e^{i k_x}-e^{i k_y}) & (e^{i k_x}+e^{i k_y})
\end{pmatrix}.
\eea
The band filling ($n$ = 1 - $x$), orbital order ($m_{ox}$ and $m_{oz}$) and charge order ($m_{c}$) parameters are determined 
self consistently by carrying out the numerical diagonalization of the Hamiltonian $\hat{H}_{CO}$ with 
\bea
n &=& \frac{1}{4N}\sum_{\k s i}(\phi^i_{s {\bf k} \gamma}\phi^{i*}_{s {\bf k} \gamma}) n(\xi^i_{\bf k}) \nonumber\\
m_{ox} &=& \frac{1}{2N}\sum_{\k i}(\phi^i_{B {\bf k} 1}\phi^{i*}_{B {\bf k} 2}+\phi^i_{B {\bf k} 2}\phi^{i*}_{B {\bf k} 1}- 
\phi^i_{D {\bf k} 1}\phi^{i*}_{D {\bf k} 2}-\phi^i_{D {\bf k} 2}\phi^{i*}_{D {\bf k} 1})n(\xi^i_{\bf k})\nonumber\\
m_{oz} &=& \frac{1}{2N}\sum_{\k i}(\phi^i_{B {\bf k} 1}\phi^{i*}_{B {\bf k} 1}-\phi^i_{B {\bf k} 2}\phi^{i*}_{B {\bf k} 2}
-\phi^i_{D {\bf k} 1}\phi^{i*}_{D {\bf k} 1}+\phi^i_{D {\bf k} 2}\phi^{i*}_{D {\bf k} 2})n(\xi^i_{\bf k})\nonumber\\
m_{c} &=& \frac{1}{2N}\sum_{\k i}(\phi^i_{B {\bf k} 1}\phi^{i*}_{B {\bf k} 1}+\phi^{i}_{B {\bf k} 2}\phi^{i*}_{B {\bf k} 2}-
\phi^i_{C {\bf k} 1}\phi^{i*}_{C {\bf k} 1}-\phi^i_{C {\bf k} 2}\phi^{i*}_{C {\bf k} 2}\nonumber\\ 
&+&\phi^i_{D {\bf k} 1}\phi^{i*}_{D {\bf k} 1}+\phi^i_{D {\bf k} 2}\phi^{i*}_{D {\bf k} 2}-
\phi^i_{A {\bf k} 1}\phi^{i*}_{A {\bf k} 1}-\phi^i_{A {\bf k} 2}\phi^{i*}_{A {\bf k} 2})n(\xi^i_{\bf k}),
\eea
where $\phi^i_{s {\bf k} \gamma}$ is the unitary matrix element from the $\gamma$-orbital of sublattice $s$ to the momentum ${\bf k}$ (in the reduced Brillouin zone) for $i$-th band. $n(\xi^i_{\k})$ is the Fermi distribution function with $\xi^i_{\k}$ = $E^i_{\k} - \mu$, where $E^i_{\k}$ is 
the $i$th eigenvalue of the Hamiltonian given by Eq. \ref{co} and $\mu$ is the chemical potential. 

\begin{figure}
\begin{center}
\vspace*{-2mm}
\hspace*{0mm}
\psfig{figure=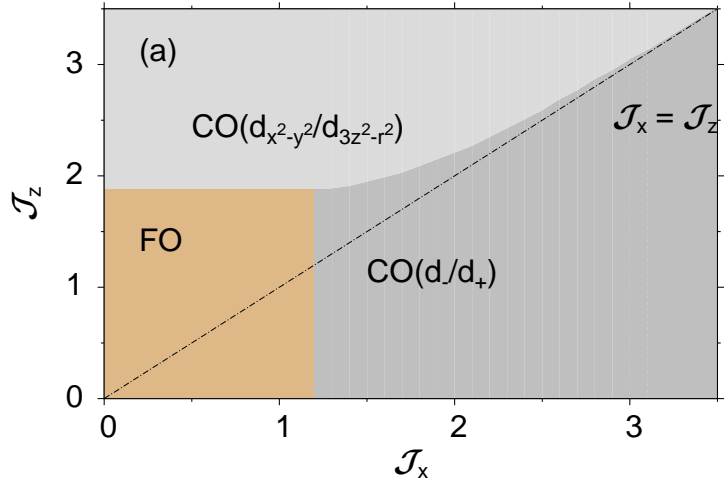,width=70mm,angle=-90}
\vspace*{-5mm}
\end{center}
\caption{$\mathcal{J}_x$ vs $\mathcal{J}_z$ phase diagram for $\varepsilon_z=0.3$, where the dashed line represents $\mathcal{J}_x$ = $\mathcal{J}_z$.}
\label{phase}
\end{figure} 
The phase diagram of $\mathcal{J}_x$-$\mathcal{J}_z$ for $\varepsilon_z=0.3$ is shown in Fig. \ref{phase}, which consists of three regions, 
ferro orbitally ordered (FO) with net positive $\langle \mathcal{T}^z_i \rangle$ orbital moment due to the anisotropy in the orbital space, CO($d_{-}/d_{+}$) with $m_{oz} = 0$, and CO($d_{x^2-y^2}/d_{3z^2-r^2}$) with $m_{ox} = 0$. The
CO($d_{-}/d_{+}$) state, which is stabilized for $\mathcal{J}_x$ $\ge$ 1.2 in the region $\mathcal{J}_x$ $\ge$  $\mathcal{J}_z$ and also in a part of region 
$\mathcal{J}_x$ $<$ $\mathcal{J}_z$, while CO($d_{x^2-y^2}/d_{3z^2-r^2}$) 
state is stabilized for $\mathcal{J}_z$ $\ge$ 1.9 in rest of the region. Each CO state has an induced charge order with momentum 2${\bf Q}$ due to the orbital ordering, where more
charges are accumulated at orbitally polarized sites. This feature is in agreement with the x-ray experiment suggesting the primary role of orbital order as the orbital correlation length was
found to be larger than the charge correlation length.\cite{wakabayashi} Further, the straight line $\mathcal{J}_x = \mathcal{J}_z$ becomes the
phase boundary line separating two CO phases in the strong coupling limit, which is due to the fact that the electron-phonon coupling term of the Hamiltonian possesses rotational
symmetry in the orbital space for $\mathcal{J}_x = \mathcal{J}_z$. If we assume that the relation $\mathcal{J}_x = \mathcal{J}_z$, which is
valid exactly only in the cubic symmetry where $e_g$ levels are degenerate, is still a good approximation in the tetragonal symmetry where two fold 
degeneracy is no longer present, then the $d_{-}/d_{+}$ orbital order is stabilized robustly except for the strong coupling regime. The charge and orbital order for the CO($d_{-}/d_{+}$) state 
is shown in Fig. \ref{ce}, where the electrons occupy $d-$ and $d_+$ orbitals at the charge accumulating bridge sites corresponding to $B$ and $D$ sublattices (see also Fig. \ref{uc}), respectively.
\begin{figure}
\begin{center}
\vspace*{-2mm}
\hspace*{0mm}
\psfig{figure=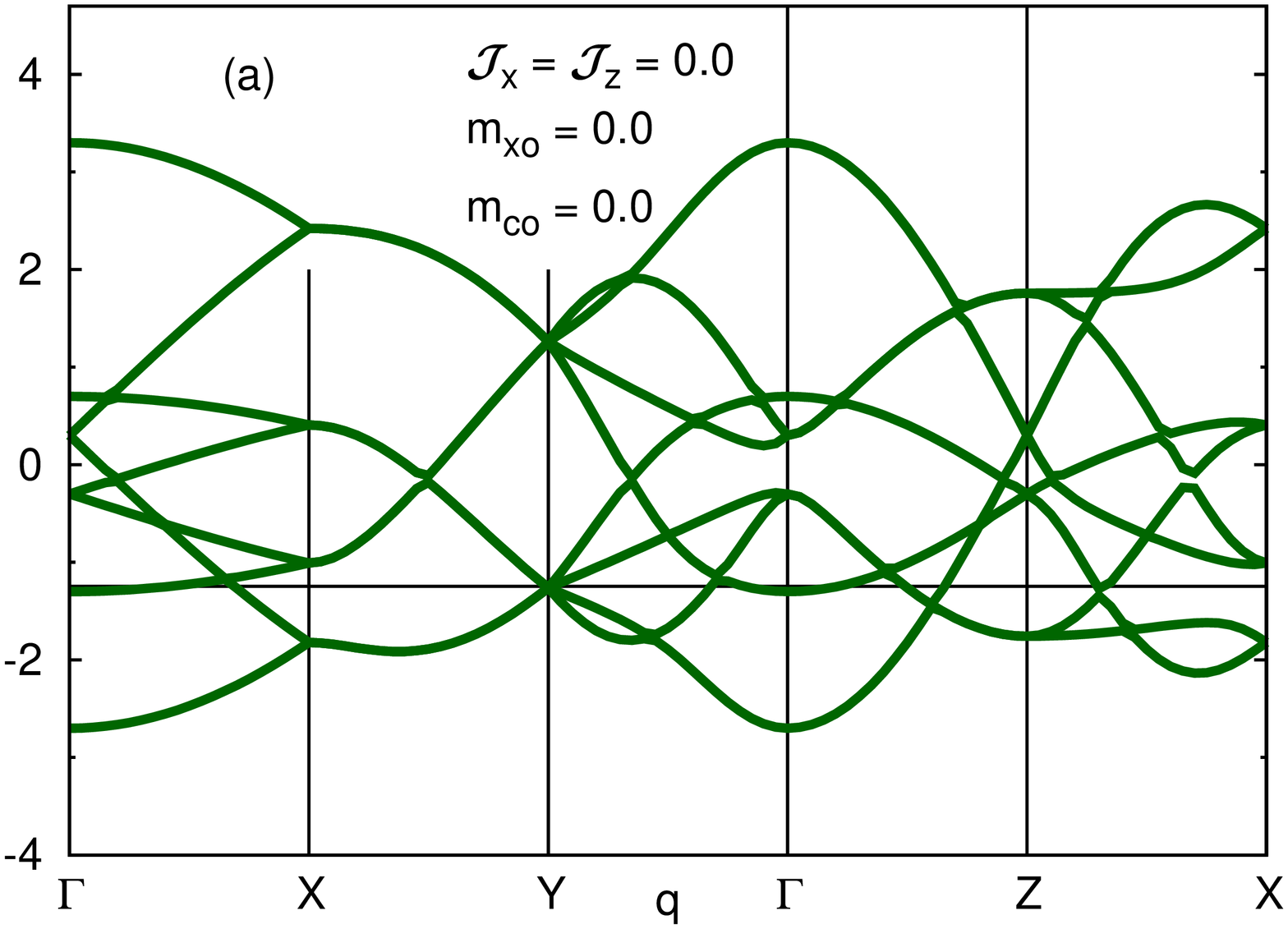,width=55mm,angle=-90}
\psfig{figure=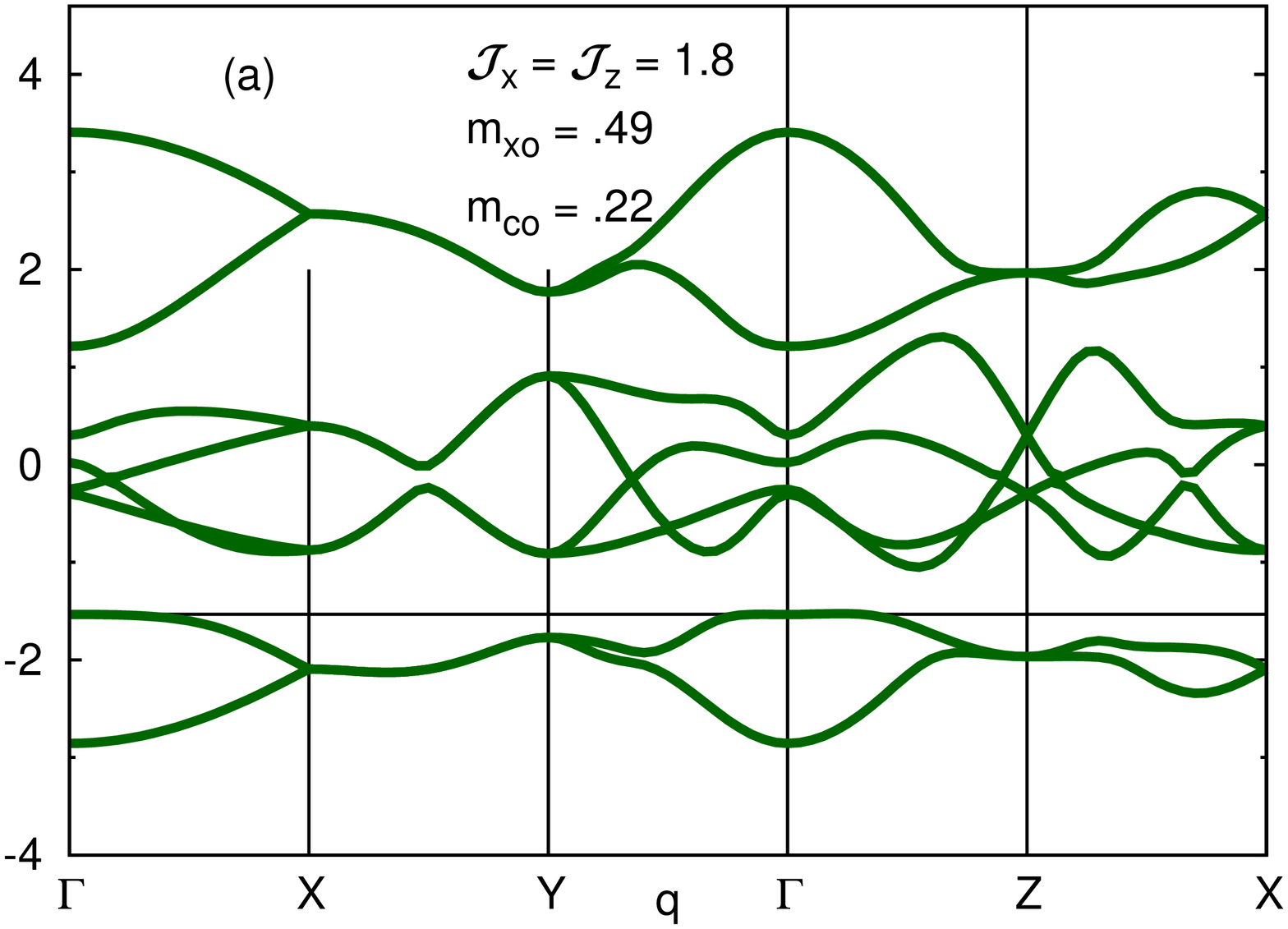,width=55mm,angle=-90}
\vspace*{-5mm}
\end{center}
\caption{Bands calculated in the sublattice basis for the CO (a) disordered and (b) ordered states along the symmetry direction of the reduced Brillouin zone with
X $(0.5\pi, 0.5\pi)$, Y $(-0.25\pi, 0.75\pi)$ and Z $(-0.5\pi, 0.5\pi)$. The CO order yields a gap opening near the Fermi surface for the hole doping at $x = 0.5$. Here, $\varepsilon_z = 0.3$ in each case.}
\label{coband}
\end{figure} 

It is important to note a crucial difference between the charge and orbital ordered states at 
$x = 0.5$ and nearby dopings. The CO state at half doping is 
a band insulator due to the gap opening between the second and third bands as shown in Fig. \ref{coband} while there will 
exist non vanishing density of states near the Fermi surface in the electron doped region $0.5 < x$. This is 
in agreement with the resistivity measurement which also exhibits a steep rise near the charge and orbital order 
transition at half doping.\cite{moritomo}

\begin{figure}
\begin{center}
\vspace*{-2mm}
\hspace*{0mm}
\psfig{figure=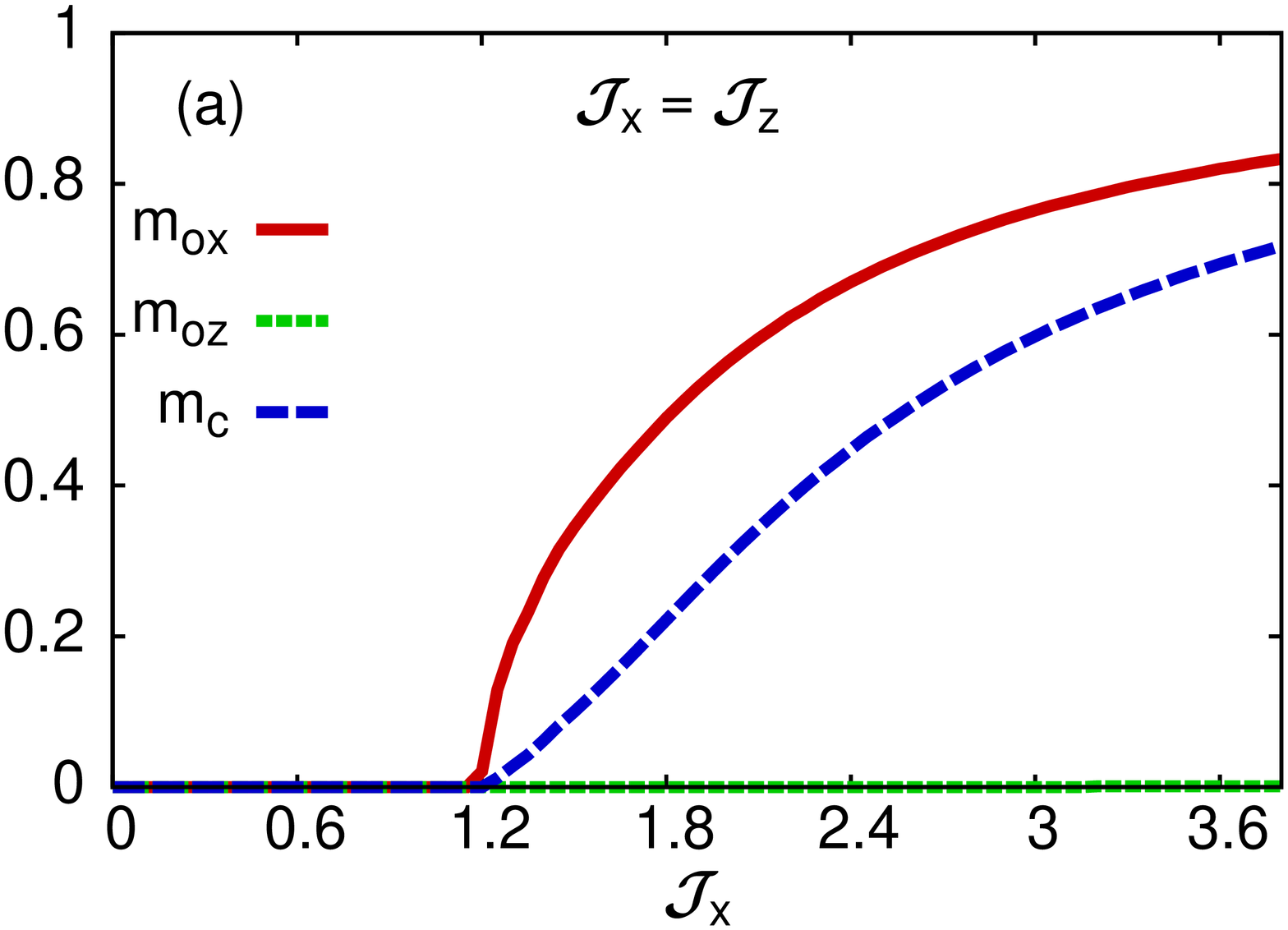,width=50mm,angle=-90}
\psfig{figure=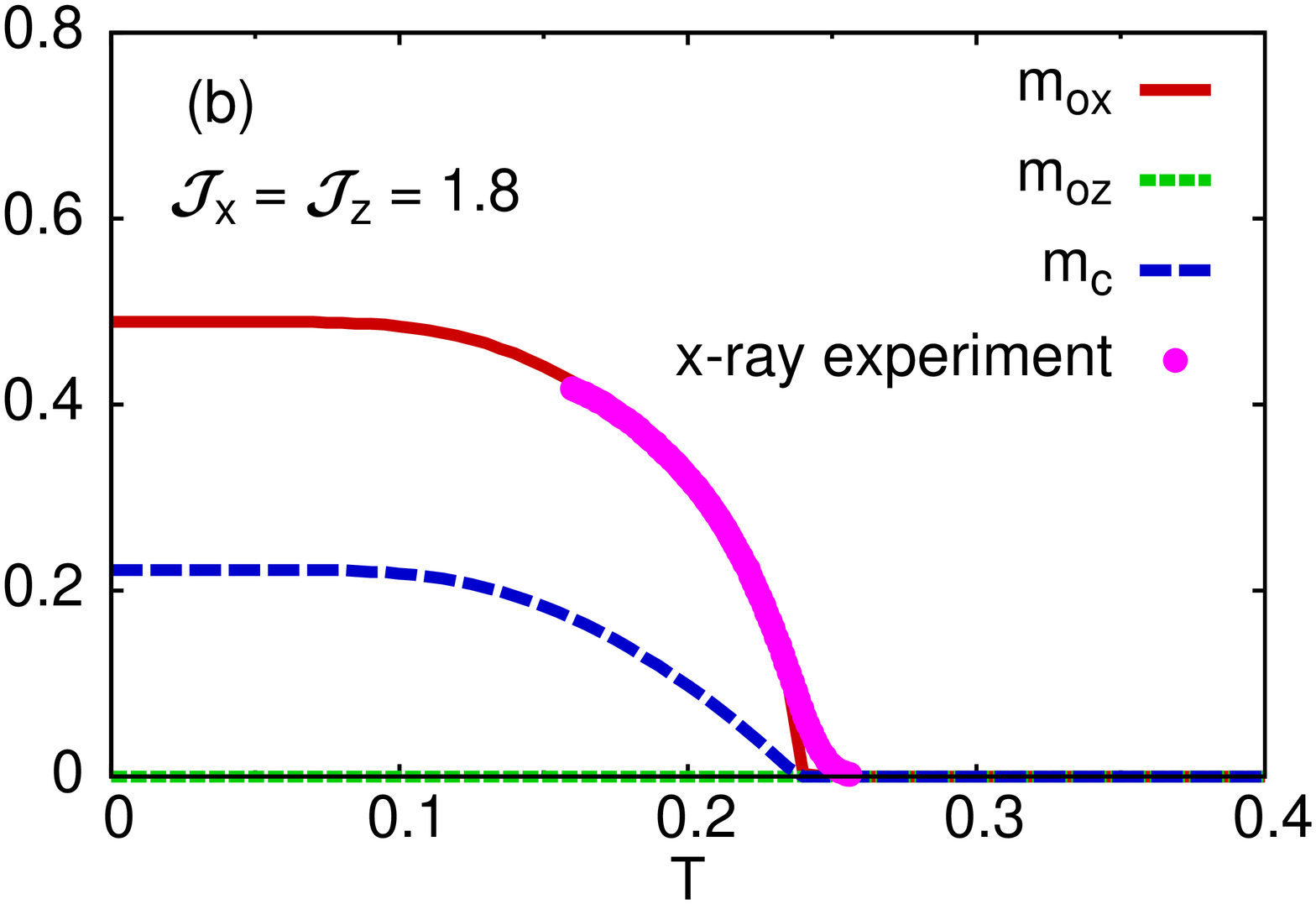,width=50mm,angle=-90}
\hspace*{2mm}
\psfig{figure=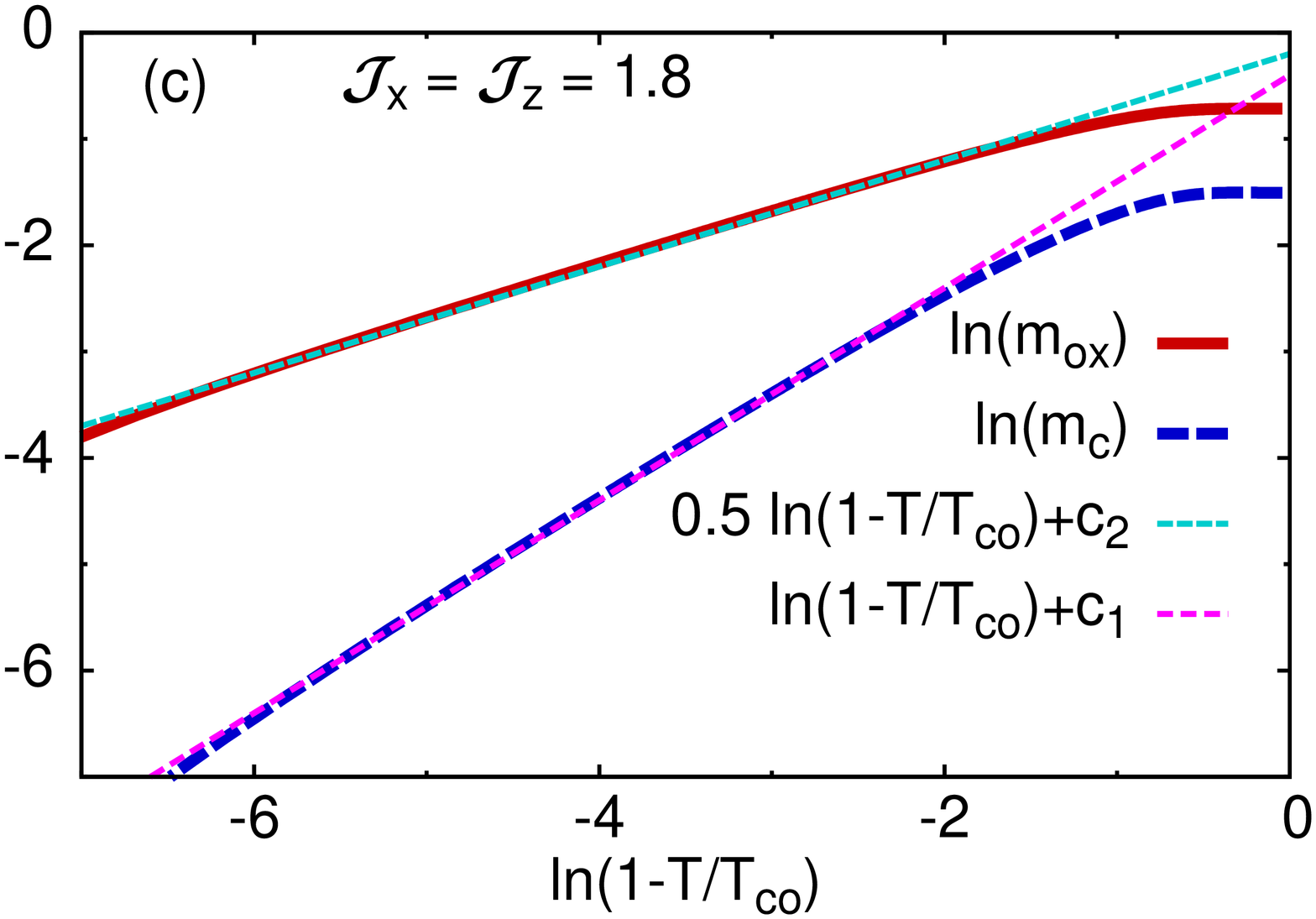,width=50mm,angle=-90}
\hspace*{0.3mm}
\psfig{figure=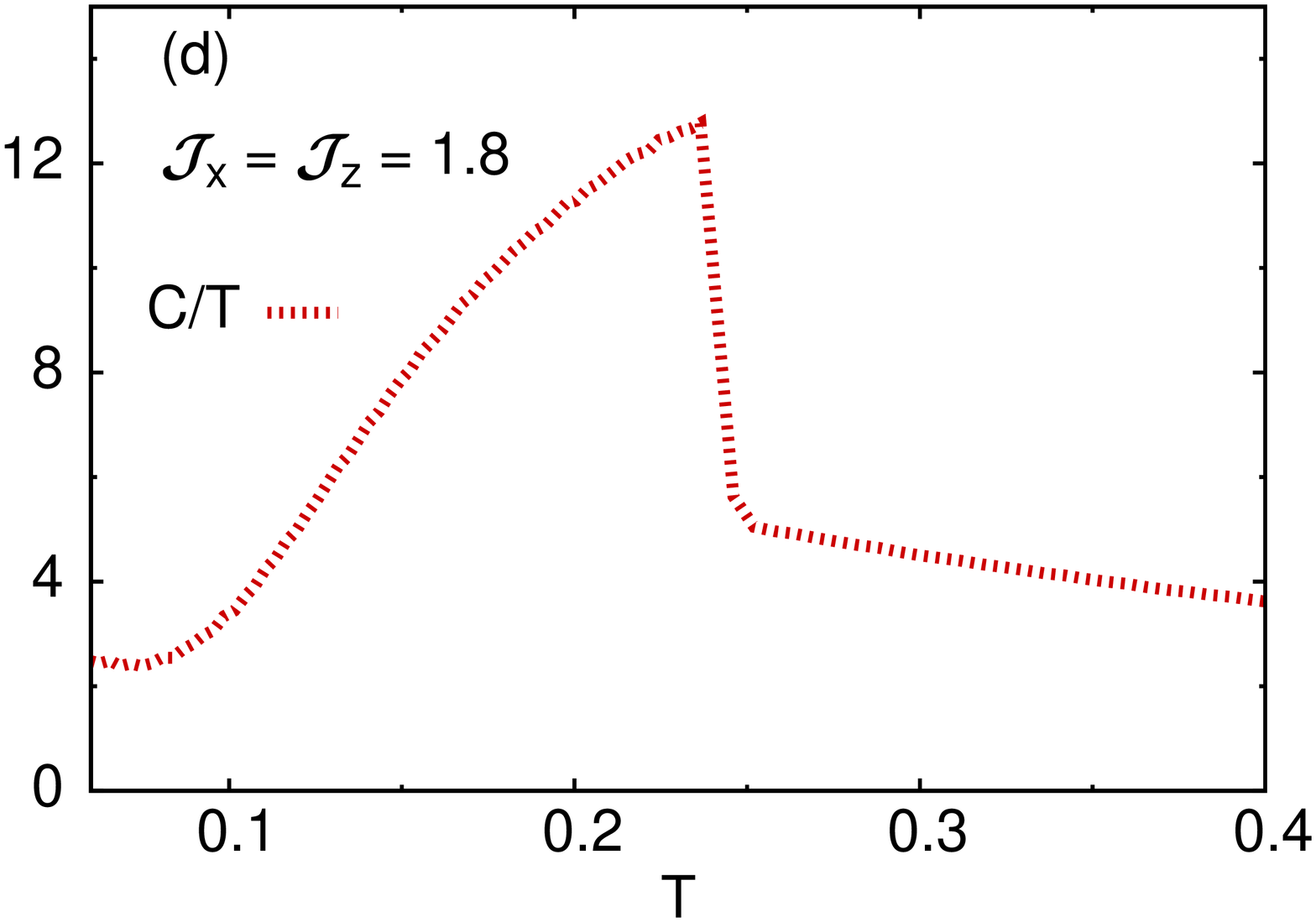,width=50mm,angle=-90}
\vspace*{-5mm}
\end{center}
\caption{(a) Orbital and charge order parameters  
as a function of interaction along $\mathcal{J}_x$ = $\mathcal{J}_z$. (b) Charge and orbital order parameter as a function of temperature 
for $\mathcal{J}_x = \mathcal{J}_z = 1.8$ with solid circles as experimental data.\cite{dhesi} (c) Order parameters vs temperature plot on the logarithmic scale near $T_{CO}$ with $c_1$ and $c_2$ as constants. (d)
Heat capacity as a function of temperature. The CEF parameter $\varepsilon_z = 0.3$.}
\label{ordp}
\end{figure} 
Fig. \ref{ordp}(a) and (b)
show the orbital and the induced charge order parameter as a function of interaction strength and temperature. Especially, the temperature dependence of the orbital order parameter is 
in good agreement with the 
resonant x-ray experiment data.\cite{dhesi} Near the transition temperature, the charge order parameter varies ($\propto T_{CO}-T$) as 
square of orbital order parameter ($\propto \sqrt{T_{CO}-T}$) as clearly shown in Fig. \ref{ordp}(c). Therefore, the nature of orbital ordering transition is second order. 
In order to check it, we also calculate the heat capacity around the transition temperature $T_{CO}$.
The heat capacity is obtained as 
\be
C= -T\frac{d^2\mathcal{F}}{dT^2},
\ee
where the free energy $\mathcal{F}$ is given by
\be
\mathcal{F}=- \sum_{\k, i} {T} \rm{ln}(1+e^{-\beta \xi^i_{\k}}) - \frac{1}{2} \sum_{l,s} \mathcal{J}_l \langle \mathcal{T}^l_s \rangle  \langle \mathcal{T}^l_s \rangle + \mu N 
\ee
with $\beta = 1/T$, while the contributions from the spin degree of freedom has not been incorporated. $\k$ is the momentum in the reduced Brillouin zone, $s$ stands for the sublattice index, and $N$ is the 
total number of electrons in a unit cell. The temperature dependence of heat capacity is shown in Fig. \ref{ordp}(d), where the heat capacity jump can be found at the transition temperature.
 
The density-wave state scenario resulting from the Fermi surface nesting can be further extended to the hole doped region
$0.5  \le  x \le 0.7 $ with the relation between orbital ordering (${\bf Q}_o$) and charge ordering wavevector (${\bf Q}_c$) as ${\bf Q}_c = 2{\bf Q}_o$ in consistent with x-ray experiments
on several single-layer manganites, where orbital ordering wavevector was observed to 
depend linearly on the hole doping
${Q}_{ox} = {Q}_{oy} = \pi (1-x)$.
\section{CE-type AFM state}
\begin{figure}
\begin{center}
\vspace*{0mm}
\hspace*{-0mm}
\psfig{figure=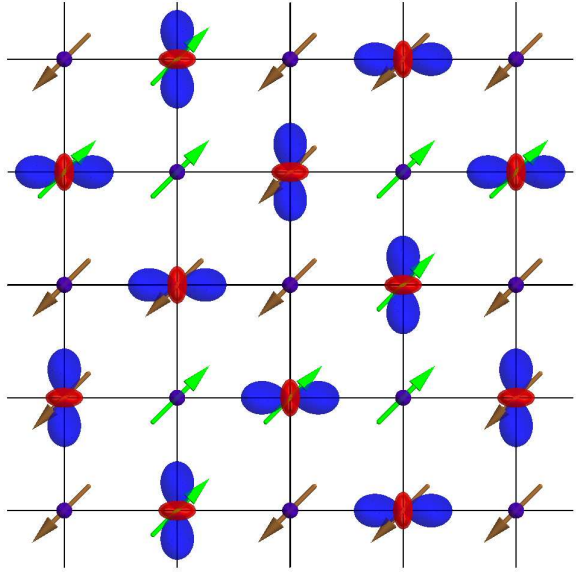,width=80mm,angle=0}
\vspace*{-5mm}
\end{center}
\caption{Orbital, charge, and spin arrangement in the CE-type state, with top view of $d_{-}$ and $d_{+}$ orbitals. Ferromagnetic
chains of opposite spin orientations have been shown in 
different colors.}
\label{ce}
\end{figure}  
Local moments, which are disordered due to the thermal fluctuations in the CO
state prior to the transition to low temperature CE-type state, try to attain a configuration with the help of the 'double exchange' mechanism\cite{zener} by which kinetic energy gain will increase in the 
system. Therefore, to find the nature of the CE state shown in Fig. \ref{ce}, we consider a meanfield Hamiltonian for an
electron with spin $\sigma$ given by 16$\times$16 matrix in a 4+4 sublattice basis
\bea
\hat{H}_{CE}({\bf k}) &=& \sum_{{\bf k}, \mu}
\Psi^{\dagger}_{\bf k}
\begin{pmatrix}
\hat{H}^{11}({\bf k}) & \hat{H}^{12}({\bf k}) \\
\hat{H}^{12 \dagger}({\bf k}) & \hat{H}^{22}({\bf k}) \\
\end{pmatrix}
\Psi_{\bf k},
\eea
where $\Psi^{\dagger}_{\bf k} = (d^{ \dagger}_{  A {\bf k} 1}, 
d^{\dagger }_{ A {\bf k} 2}..d^{\dagger }_{ D {\bf k} 1}, d^{ \dagger}_{ D {\bf k} 2},d^{ \dagger}_{A^{\prime} {\bf k} 1}, 
d^{\dagger}_{ A^{\prime} {\bf k} 2}... d^{ \dagger}_{ D^{\prime} {\bf k} 1} ,d^{ \dagger}_{ D^{\prime} {\bf k} 2} )$ is the electron field operator
with primed $d$ electron 
operators acting on the basis with spin oriented in the opposite directions. We note that $A$, $B$, $C$, and $D$ are the same sublattice indices considered for the CO state with 
$B$ and $D$ sites having more electron density in $d_-$- and $d_+$-orbitals, respectively , while $A$ and $C$ are charge deficient. Sublattices $A^{\prime}$, $B^{\prime}$, $C^{\prime}$, and $D^{\prime}$ have similar charge and orbital structure as the 
sublattices $A$, $B$, $C$, and $D$, respectively. Therefore, while $\hat{H}^{11}({\bf k})$ and $\hat{H}^{22}({\bf k})$ describes charge and orbital structure on up- and down-spin chains, the elements of $\hat{H}^{12}({\bf k})$ are the 
Bloch phase factors connecting these two different set of sublattices. These block matrices are given by
\bea 
\hat{H}^{11(22)}({\bf k}) = 
\begin{pmatrix}
\hat{\Delta}_{cr} & \hat{K}_x & \hat{0} & \hat{K}^{\dagger}_y \\
\hat{K}^{\dagger}_x & -\hat{\Delta}_o+\hat{\Delta}_{cr} & \hat{K}_x & \hat{0} \\
\hat{0} & \hat{K}^{\dagger}_x & \hat{\Delta}_{cr} & \hat{K}_y \\
\hat{K}_y & \hat{0} & \hat{K}^{\dagger}_y & \hat{\Delta}_o+\hat{\Delta}_{cr}
\end{pmatrix}
\mp \sigma J_HS\hat{{\bf 1}},
\eea
and 
\bea
\hat{H}^{12}({\bf k}) = 
\begin{pmatrix}
\hat{0} & \hat{K}_y & \hat{0} & \hat{K}^{\dagger}_x \\
\hat{K}^{\dagger}_y & \hat{0} & \hat{K}_y & \hat{0} \\
\hat{0} & \hat{K}^{\dagger}_y & \hat{0} & \hat{K}_x \\
\hat{K}_x & \hat{0} & \hat{K}^{\dagger}_x & \hat{0} 
\end{pmatrix},
\eea
where $\hat{K}_x$ and $\hat{K}_y$ are matrices of Bloch phase factors in the $x$- and $y$-directions described by Eq. 13, and $\hat{{\bf 1}}$ is a 8$\times$8 identity matrix. The band filling, orbital and 
charge order parameters are determined self consistently as in the CO state by 
numerically diagonalizing the Hamiltonian $\hat{H}_{CE}$.
\begin{figure}
\begin{center}
\vspace*{-2mm}
\hspace*{0mm}
\psfig{figure=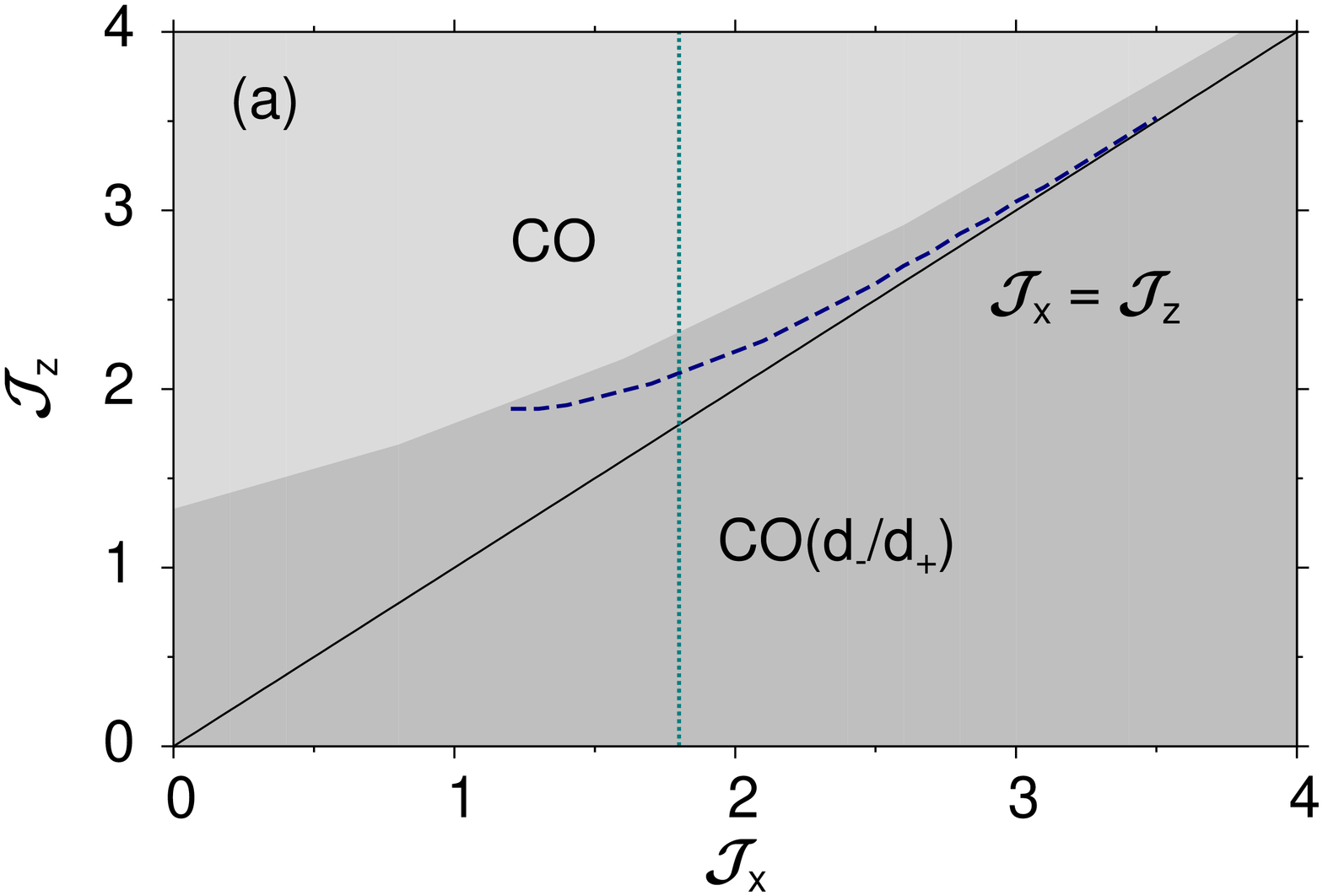,width=65mm,angle=-90}
\psfig{figure=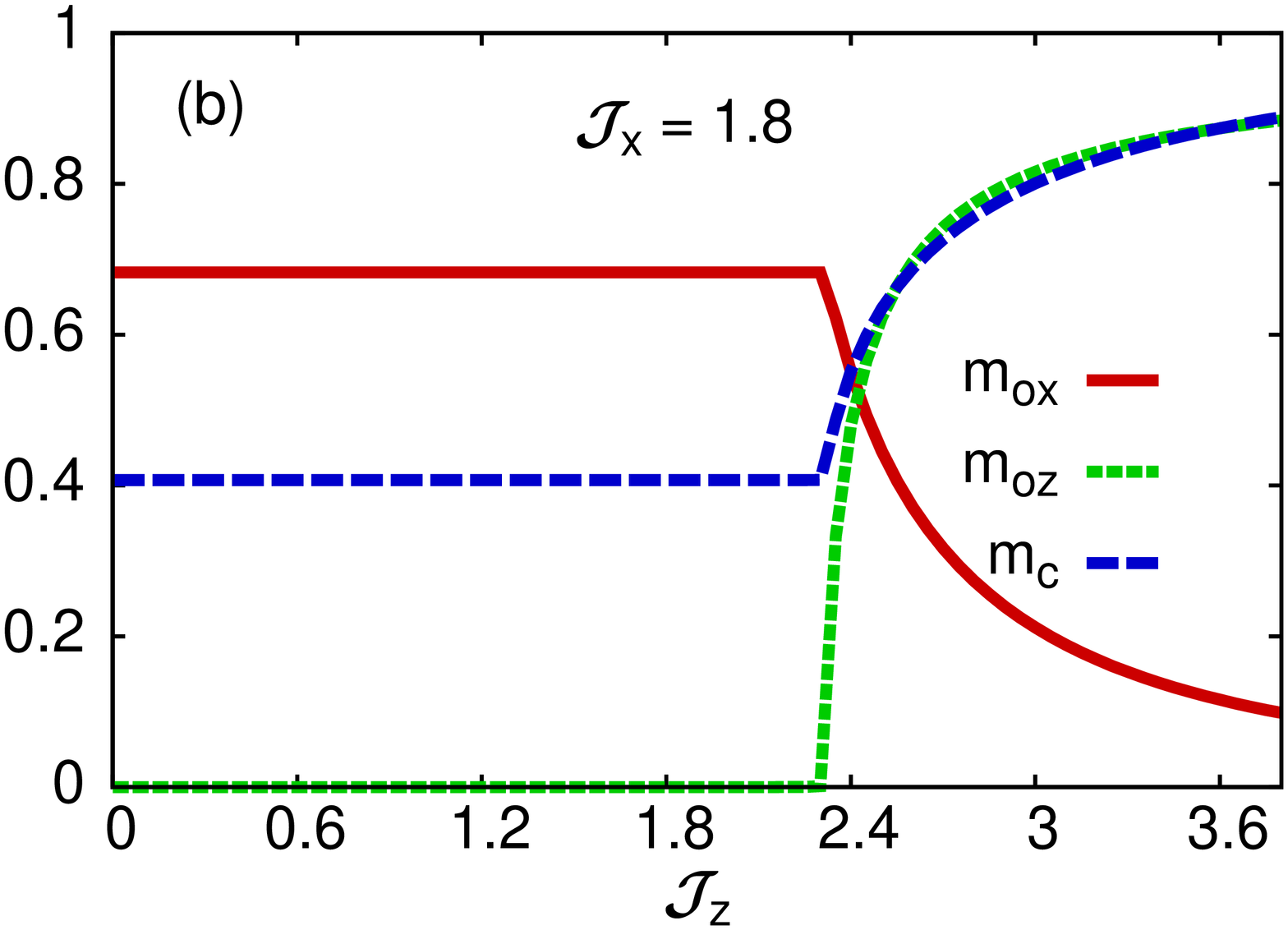,width=55mm,angle=-90}
\psfig{figure=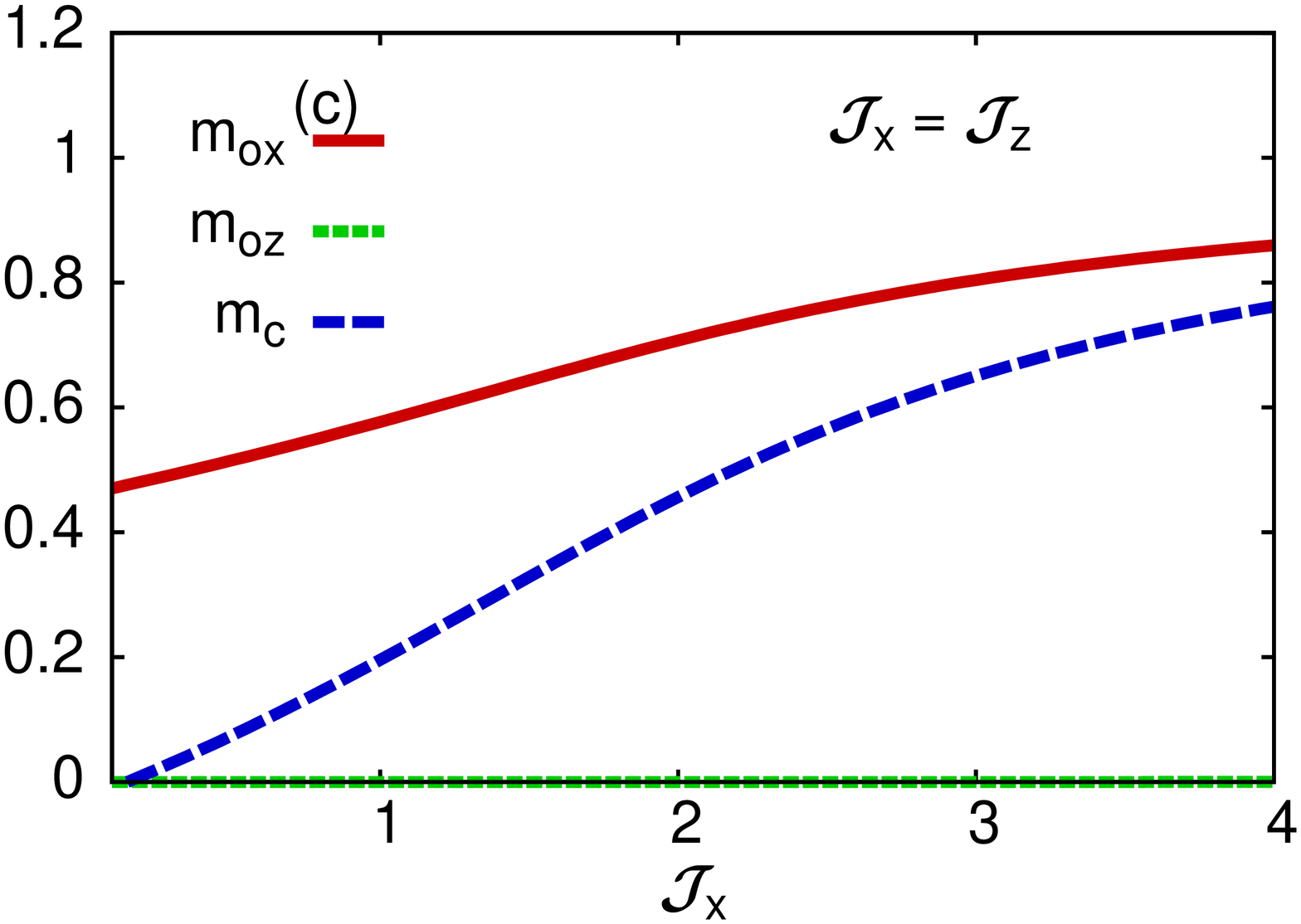,width=55mm,angle=-90}
\vspace*{-5mm}
\end{center}
\caption{In the CE-type spin arrangement with  $J_H = 8t$ and $\varepsilon_z = 0.3$ (a) $\mathcal{J}_x$ vs $\mathcal{J}_z$ phase diagram, where dotted line corresponds to $\mathcal{J}_x$ = 1.8, and dashed line represents the 
phase-boundary line between CO$(d_{x^2-y^2}/d_{3z^2-r^2})$ and CO$(d_-/d_+)$ of Fig. \ref{phase}, (b) nature of orbital order parameter in 
the CO state as a function of $\mathcal{J}_z$ for fixed $\mathcal{J}_x$ = 1.8, (c) orbital order ($m_{ox}$ and $m_{oz}$) and charge order ($m_{c}$) parameters as a function of interaction $\mathcal{J}_x$, where
$\mathcal{J}_x = \mathcal{J}_z$.}
\label{cephs}
\end{figure} 
 
Fig. \ref{cephs}(a) shows the phase diagram of $\mathcal{J}_x$ vs $\mathcal{J}_z$ with $J_H = 8t$ in the CE-type spin arrangement, which 
consists of two types of charge and orbitally ordered state, CO($d_-/d_+$) stabilized for $\mathcal{J}_x$ $\ge$ $\mathcal{J}_z$ and 
also in a part of the region $\mathcal{J}_x$ $<$ $\mathcal{J}_z$, another CO state with the character of orbital ordering depending 
on the strength of $\mathcal{J}_z$ stabilized in the rest of the region. The latter CO state transforms to the CO($d_{x^2-y^2}/d_{3z^2-r^2}$)
in the limit of large $\mathcal{J}_z$ as shown in Fig. \ref{cephs}(b). Here, the phase boundary line is pushed further inside the 
region $\mathcal{J}_z$ $ > $ $\mathcal{J}_x$ as compared to 
the CO state without CE-type spin arrangement. In other words, CE state supports in breaking the four-fold rotational symmetry
in the orbital space due to the zig-zag spin arrangement with the help of double-exchange mechanism, and will stabilize the 
$d_{-}/d_{+}$ orbital order further. This is further highlighted by Fig. \ref{cephs}(c) through the dependence of orbital order parameter on 
the interaction in the CE-type spin arrangement, where a near-complete orbital polarization 
with non-zero $m_{ox}$ exists even in limit of vanishing interaction strength at sites like $B$ and $D$. Moreover, we notice a significant increase in the magnitude of both charge ($\approx$ 100 $\%$ ) and orbital order ($\approx$ 40 $\%$ ) 
parameters for the same value of $\mathcal{J}_x$ =  $\mathcal{J}_z$ = 1.8 for the CE state as compared to the CO state without any spin order. A jump in the orbital order parameter near the transition from the CO to the CE state 
has also been noticed in the x-ray experiments.\cite{wilkins}
 
\begin{figure}
\begin{center}
\vspace*{-2mm}
\hspace*{0mm}
\psfig{figure=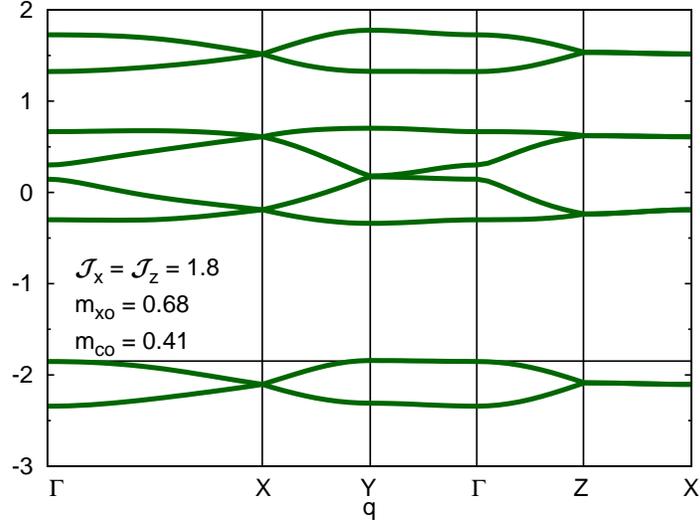,width=70mm,angle=-90}
\vspace*{-5mm}
\end{center}
\caption{Eight lower subbands calculated for the CE-state along the symmetry direction of the reduced Brillouin zone with
X $(0.5\pi, 0)$, Y $(0.25\pi, -0.25\pi)$ and Z $(0.25\pi, 0.25\pi)$.  There exists a gap near the Fermi surface between the second and third band 
for the hole doping at $x = 0.5$ as in the case of CO state. The value of CEF parameter $\varepsilon_z = 0.3$.}
\label{ce_band}
\end{figure} 
Fig. \ref{ce_band} shows eight lower sub-bands while other eight subbands are shifted by $\approx 2JS$. It is interesting to note that
the CE-state is a band insulator like the CO state due to the gap near the Fermi surface at $x = 0.5$. However, an important difference is that the electrons can hop along the zigzag chain in the CE-state 
unlike the CO state. In this spin arrangement, the electrons avoid hopping
in the direction perpendicular to the chain due to less 
kinetic energy gain as the hopping parameter is very small 
in the perpendicular direction, which will force the neighboring 
zigzag chains to have spins pointing in the opposite directions through the super-exchange process. 
However, this fine balance
in cooperation between orbital density-wave and a commensurate spin ordering
due to the large $t_{2g}$ local moments, which exists only near half doping, may be destroyed on moving away from the half doping.

The stabilization of CO($d_-/d_+$) for $\mathcal{J}_x$ $\approx$  $\mathcal{J}_z$ even in the CE-type spin arrangement is not surprising if we look at 
the hopping matrix in the new basis consisting of $d_+$ and $d_{-}$ orbitals  
\bea
-\frac{1}{2}
\begin{pmatrix}
a c_x + b c_y & c_x+c_y \\
 c_x+c_y & b c_x+a c_y
\end{pmatrix},
\eea
where $c_x = \cos k_x$,  $c_y = \cos k_y$, $a = 2-\sqrt{3}$, $b = 2+\sqrt{3}$. 
In this new basis, the ratios of hopping parameters in the $x$- and $y$-directions for the $d_{-}$ and $d_{+}$ orbitals
are $t^{}_{- - y}$/$t^{}_{- - x}$ =$t^{}_{+ + x}$/$t^{}_{+ + y}$ $\approx$ 
0.07. Thus, the intraorbital hopping is quasi-one dimensional for each orbital, that is, the dominant hopping 
occurs in the $x$- and $y$-directions for the $d_-$ and $d_+$ orbitals, respectively. At the same time, interorbital hopping is  
isotropic with respect to the magnitude and phase. Therefore in the CO ($d_{-}/d_{+}$) state, kinetic energy gain can be maximized through the double-exchange mechanism 
if the electrons move along the zigzag chains as shown in Fig. \ref{ce}.
In other words, the electrons hop through $d_{-}$ and $d_{+}$ orbitals along $x$- and $y$- directions at bridge sites like $B$ and $D$, 
respectively, while the corner sites like $A$ and $C$ will have equal mixture of $d_{-}$ and $d_{+}$ orbitals as 
interorbital hopping is isotropic and equal for both the orbitals. This type of orbital order may also play an important role stabilizing the CE-type state previously obtained in the psuedo-cubic 
manganites using Monte Carlo simulation.\cite{yunoki}
\section{Conclusions}
In conclusion, we have investigated the nature of orbital order in a two orbital tightbinding model for half-doped layered La$_{0.5}$Sr$_{1.5}$MnO$_4$ with realistic 
electronic state as observed in the ARPES measurement. The orbital order of $B_{1g}$-representation with $\Q = (0.5\pi, 0.5\pi)$, which results from the 
strong nesting between the portions of the Fermi surface having $d_{\pm}$ orbital character predominantly, induces the charge order parameter with $2\Q$, whose 
magnitude develops according to $T_{c}-T$ by decreasing temperature around the transition point. The temperature dependence of orbital-order parameter agrees well 
with the x-ray experiments. For half doping, the charge and orbital ordered state is 
a band insulator in accordance with the resistivity measurement which displays steep rise near the transition. Furthermore, it has been shown that the CE-type spin 
arrangement stabilizes the same orbital order even in the absence of the orbital exchange interaction, which is reflected in the jump
of the orbital order parameter. Band structure in the CE-type state suggests that it is also a band insulator like the 
charge and orbitally ordered state without any spin order.
\section*{Appendix}
The meanfield Hamiltonian corresponding to Eq. \ref{green} in the momentum basis $\k, \k+\Q, \k+2\Q$ and $\k+3\Q$ is given by   
\bea
\hat{\mathcal{H}}_{CO}({\bf k}) &=& \sum_{{\bf k}, \mu}
\Phi^{\dagger}_{\bf k}
\begin{pmatrix}
\hat{K}_{\k}+\hat{\Delta}_{cr} & -\hat{\Delta}_o & \hat{0} & -\hat{\Delta}_o \\
-\hat{\Delta}_o & \hat{K}_{\k+\Q} +\hat{\Delta}_{cr} & -\hat{\Delta}_o & \hat{0} \\
\hat{0} & -\hat{\Delta}_o & \hat{K}_{\k+2\Q} + \hat{\Delta}_{cr} & -\hat{\Delta}_o \\
-\hat{\Delta}_o & \hat{0} & -\hat{\Delta}_o & \hat{K}_{\k+3\Q}+\hat{\Delta}_{cr}
\end{pmatrix}
\Phi_{\bf k},
\eea
where $\Phi^{\dagger}_{\bf k} = (d^{\dagger}_{ {\bf k} 1}, 
d^{\dagger}_{ {\bf k} 2}, d^{\dagger}_{ {\bf k+Q} 1}, d^{\dagger}_{ {\bf k+Q} 2} ...d^{\dagger}_{ {\bf k+3Q} 1}, d^{\dagger}_{{\bf k+3Q} 2} )$ and
\be
\hat{K}_{\k} =
\begin{pmatrix}
-\frac{3}{2}(\cos k_x + \cos k_y) &  \frac{\sqrt{3}}{2}(\cos k_x - \cos k_y) \\
 \frac{\sqrt{3}}{2}(\cos k_x - \cos k_y) & -\frac{1}{2}(\cos k_x + \cos k_y)
\end{pmatrix}.
\ee
This Hamiltonian can then be transformed to the Hamiltonian in the sublattice basis (Eq. \ref{co}) using following unitary transformation  
\bea
\hat{U} &=& 
\begin{pmatrix}
\hat{\tau}^0 & \hat{\tau}^0 & \hat{\tau}^0 & \hat{\tau}^0 \\
\hat{\tau}^0 & i\hat{\tau}^0 & \hat{\tau}^0 & -i\hat{\tau}^0 \\
\hat{\tau}^0 & -\hat{\tau}^0 & \hat{\tau}^0 & -\hat{\tau}^0 \\
\hat{\tau}^0 & -i\hat{\tau}^0 & \hat{\tau}^0 & i \hat{\tau}^0
\end{pmatrix}. 
\eea
\section* {Acknowledgements} 
This work is supported by Basic Science Program through the National Research Foundation of Korea (NRF) funded by the Ministry of Education (NRF-2012R1A1A2008559). 
D. K. Singh would also like to acknowledge the Korea Ministry of Education, Science and Technology, Gyeongsangbuk-Do and Pohang City for the support of the Young Scientist Training program at the
Asia-Pacific Center for Theoretical Physics. The authors thank K. H. Lee for useful discussions.

\end{document}